\documentclass[aps,pra,superscriptaddress,twocolumn]{revtex4}
\usepackage{bm} \usepackage{graphicx} \usepackage{amsmath}
\usepackage{amssymb} 
\usepackage{braket,mleftright}
\usepackage{fancyhdr}
\usepackage{float}

\begin{document}

\title{Free Nano-Object Ramsey Interferometry for Large Quantum Superpositions}

\author{C. Wan}
\affiliation{QOLS, Blackett Laboratory, Imperial College London, London SW7 2BW, United Kingdom}

\author{M. Scala}
\affiliation{QOLS, Blackett Laboratory, Imperial College London, London SW7 2BW, United Kingdom}

\author{G. W. Morley}

\affiliation{Department of Physics, University of Warwick, Gibbet Hill Road, Coventry CV4 7AL, United Kingdom}

\author{ATM. A. Rahman}
\affiliation{Department of Physics, University of Warwick, Gibbet Hill Road, Coventry CV4 7AL, United Kingdom}
\affiliation{Department of Physics and Astronomy, University College
London, Gower St., London WC1E 6BT, United Kingdom}

\author{\mbox{H. Ulbricht}}
\affiliation{Department of Physics and Astronomy, University of Southampton, Southampton, SO17 1BJ, United Kingdom}

\author{J. Bateman}
\affiliation{Department of Physics, College of Science, Swansea University, Swansea SA2 8PP, UK}

\author{P. F. Barker}
\affiliation{Department of Physics and Astronomy, University College
London, Gower St., London WC1E 6BT, United Kingdom}

\author{S. Bose}\email{s.bose@ucl.ac.uk}\thanks{Corresponding Author}
\affiliation{Department of Physics and Astronomy, University College
London, Gower St., London WC1E 6BT, United Kingdom}

\author{M. S. Kim}
\affiliation{QOLS, Blackett Laboratory, Imperial College London, London SW7 2BW, United Kingdom}
\date{\today}
\begin{abstract}
We propose an interferometric scheme based on an untrapped nano-object subjected to gravity. The motion of the center of mass (c.m.) of the free object is coupled to its internal spin system magnetically, and a free flight scheme is developed based on coherent spin control. The wavepacket of the test object, under a spin-dependent force, may then be delocalized to a macroscopic scale. A gravity induced dynamical phase (accrued solely on the spin state, and measured through a Ramsey scheme) is used to reveal the above spatially delocalised superposition of the spin-nano-object composite system that arises during our scheme. We find a remarkable immunity to the motional noise in the c.m. (initially in a thermal state with moderate cooling), and also a dynamical decoupling nature of the scheme itself. Together they secure a high visibility of the resulting Ramsey fringes. The mass independence of our scheme makes it viable for a nano-object selected from an ensemble with a high mass variability. Given these advantages, a quantum superposition with $100$ nm spatial separation for a massive object of $10^9$ amu is achievable experimentally, providing a route to test postulated modifications of quantum theory such as continuous spontaneous localisation.

\end{abstract}

\maketitle

{\em Introduction.---}It is expected by a significant community of researchers that when one reaches a superposition of quantum states separated spatially by $\sim 100$ nm for objects of mass $\sim10^9$ amu or larger, some hitherto unseen modifications of quantum mechanics \cite{bassi2013models,Adler2007} or self gravitational effects (Schr\"odinger-Newton equations) \cite{BassiSNE} may start manifesting. Even practically, such highly non-classical states will have varied applications in quantum technology such as in metrology. Hence generating such states, and indeed evidencing them, is of prime importance in the macroscopic frontier of quantum technology. Over the years several proposals for probing spatial superpositions of confined macroscopic objects have been proposed \cite{SBosePRA1999, Armour-Blencowe-Schwab-2002, marshall2003towards, RablPRB2009, romero2010toward, barker2010cavity, chang2010cavity, gerlich2011quantum, romero2012quantum, Twamley2012, yin2013large}, but tethering/trapping naturally limits the distance that superposed state can be separated, and the trapping mechanism itself might offer a route to decoherence. Thus many recent proposals involve free flight -- they have proposed to achieve large spatial superpositions through nonlinear optomechanics using cavity induced measurements \cite{romero2011large, Oril-PRA} and through Talbot intereference of a nano-object ensemble \cite{JamesNatureCom}. However, access to strong optomechanical nonlinearities and/or the conditional preparation of superpositions are required in the former set of proposals, while mass dispersion is a difficulty encountered in the latter type of proposals. Here we thus propose to use Ramsey interferometry of untrapped nano-objects to create and probe superpositions. The scale of the superposition is controllable through flight time and magnetic field gradients, while the mass does not appear in the relevant interferometric phase.

In this Letter, we propose a scheme based on a free, thermal nano-object with the motion of the centre of mass (c.m.) coupled to its internal state. Under coherent control on the internal state 
the wavepacket of the particle could be split and merged in a double-slit interferometry fashion. 
If further, the interferometric arms are subjected to different gravitational potentials, a dynamical phase is induced (just as with the neutron interferometry experiments of Ref.\cite{Collela-Overhauser-Werner}) and measured solely on spin state which evidences the spatially separated superposition of the test object. 
The phase itself is independent of the mass so that the nanoparticle ensemble used in the experiment can have a wide range of masses of about the same order of $10^9$ amu. With the capability of generating a highly spatially separated superposition and being robust to motional noise,  our system paves the way to testing some modifications of quantum theory, such as continuous spontaneous localisation (CSL) \cite{CSLPearle1989, NonmarkovianCSL, AlderCSL}. 

\begin{figure}[h]
\centering
\includegraphics[width=5cm]{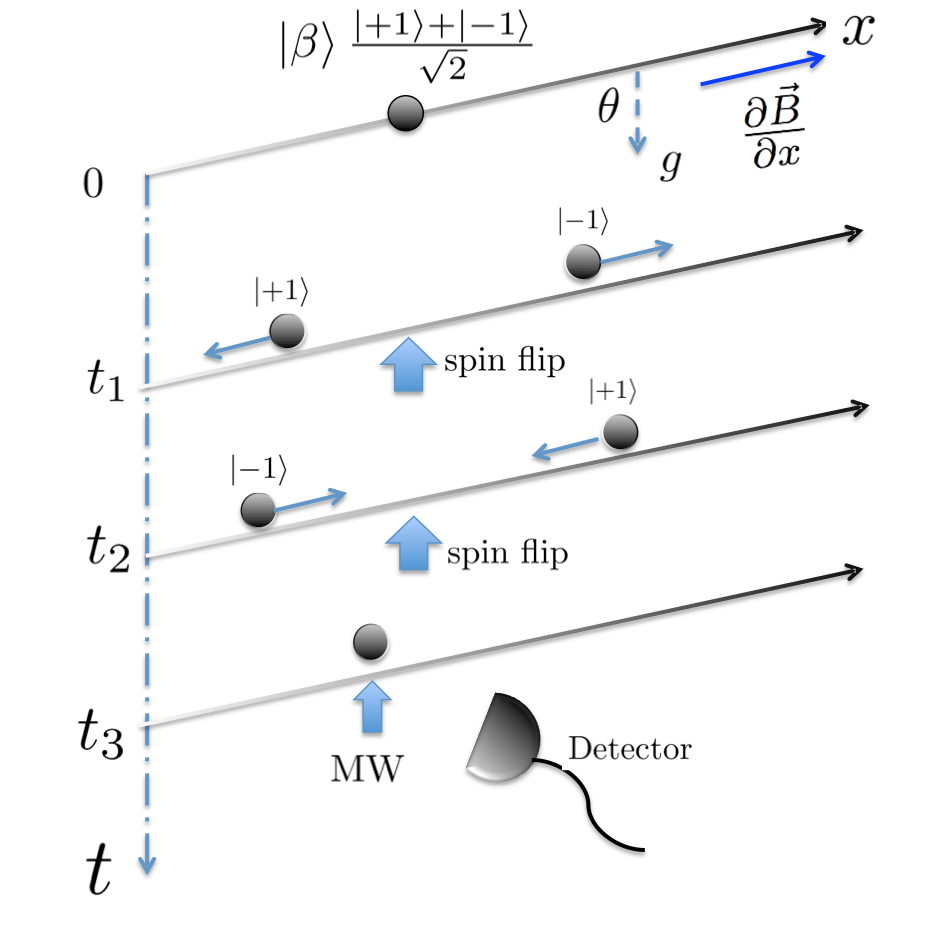}
\caption{An untrapped nano-object undergoes an illustrated interferometric scheme. A magnetic field gradient (titled by $\theta$ with respect to gravity) couples the c.m. and the spin of the particle. 
Starting with a spin state $(\Ket{+1} + \Ket{-1})/\sqrt{2}$ at $t=0$, the wave packets of the particle split and accelerate until time $ t_1$,  when a set of microwave (MW) pulse is sent to flip the spin states, which decelerates both the wave packet components leading to their motion along the axis reversing after a relevant time. The second set of MW pulses, sent at time $t_2$, reverses the direction of acceleration of the separated wave-packet components once again so that after $ t_2$ they start to decelerate while approaching each other and merger together at $t_3$, when a MW pulse is sent to perform the Ramsey measurement.}
\label{FFS}
\end{figure}
 {\it Model.---}As shown in Fig. \ref{FFS}, we first assume that a nano-diamond with a single spin-1 nitrogen-vacancy (NV) center is prepared with its c.m. in a low temperature thermal state in a harmonic trap, say, by feedback cooling \cite{feedbackCooling, CoolingLevitatedObject}. The NV spin's symmetry axis is aligned with the trapping axis, $x$, and with its spin state initialised to $\Ket{0}$ (by standard optical pumping). The trapping axis $x$ is tilted by $\theta$ with respect to the direction of the gravitational field and after that a uniform magnetic field gradient $ \partial \vec{B} /\partial x$ is introduced which covers a certain region in the vicinity of the trapped particle and couples its spin and motional degrees of freedom (DOF) along $x$.
 
 
Starting at $t=0$ we release the nano-object and immediately send a microwave (MW) pulse that creates a spin superposition $(\Ket{+1} + \Ket{-1})/\sqrt{2}$. The untrapped particle will propagate freely under a spin dependent force and the gravity of the mass, the corresponding Hamiltonian is
\begin{equation}
H = \frac{\hat{p}^2}{2m} - g_{NV}\mu_B \frac{\partial B}{\partial x} \hat{S_z} \hat{x} + m g\cos{\theta} \hat{x},
\label{Hamil_sys}
\end{equation}
where $\mu_B$ is the Bohr magneton, $g_{NV}$ is the Lande g-factor, $\theta$ the tilting angle of the initial trap with respect to the gravitational direction, $g$ the free fall acceleration, $\hat{S_z}$ is the spin $z$ operator of the NV spin,  $\hat{p}$ and $\hat{x}$ are the momentum and position operator along the trapping axis respectively. We consider the c.m. initially to be an {\em arbitrary} coherent state $\Ket{\beta}$, under Hamiltonian (\ref{Hamil_sys}) the particle will propagate in a way that its wave packets spatially separate and accelerate along $x$. The state at time $t$ is then:
\begin{equation} 
\Ket{\Psi(t)} = \frac{\Ket{\psi(t, +1)}\Ket{+1} + \Ket{\psi(t, -1)}\Ket{-1}}{\sqrt{2}},
\end{equation} 
which is the superposition we aim to demonstrate by the following Ramsey scheme. We flip  the spin state of each counter-propagating component at some appropriate times $t_1$ and $t_2$, by which the split wave packets would merge back after a relevant time, forming a two arm interferometer. The spin flip operation (from $\Ket{+1}$ to $\Ket{-1}$ or the other way) could be achieved via a two-MW-pulses sequence, provided that the Zeeman splitting due to local magnetic field is comparably large with respect to the MW pulse bandwidth \cite{supplementary}. 
If the timing of the spin manipulation is controlled by $t_1 = \frac{1}{3}t_2 =\frac{1}{4}t_3$ we would obtain a separable state at time $t_3$. Temporally the MW pulse timing is precise to 5 ns or better and the duration of each pulse would be as short as 10 ns. While the uncertainty in pulse sequences would in principle result in decoherence on the reduced spin state at the end, however such an effect would be negligible if the total free flight time is much larger than the pulse times.. The state at $t_3$ is given by 

\begin{equation} 
\Ket{\Psi(t_3)} = \frac{1}{\sqrt{2}}\Ket{\psi(t_3)}(\Ket{+1} + e^{-i\phi_g }\Ket{-1}),
\end{equation} 
where $\Ket{\psi(t_3)}$ is the final motional state of the c.m., written in position representation as
\begin{equation}
\begin{aligned}
\Braket{x|\psi(t_3)} = e^{-ip_0x}e^{-\frac{(x-x_0-p_0t_3/m-g\cos{\theta}t_3^2/2)^2}{2(\sigma^{\prime})^2}},
\end{aligned}
\label{gaussian}
\end{equation}
where $p_0$ and $x_0$ are the initial momentum and position of the nano object respectively, and $\sigma^{\prime}$ is the wave packet spread at time $t_3$ \cite{supplementary}. By dropping a global phase factor, we have 
$
\phi_g = \frac{1}{16\hbar}gt_3 ^3 g_{NV}\mu_B \frac{\partial B}{\partial x}\cos{\theta},  
$
which is the extra phase stemming from the superposition of spatially separated trajectories subjected to an auxiliary field (local gravity in this case). It could be measured by completing the Ramsey scheme: The second MW pulse on the NV spin at time $t_3$ will map this phase to the population of state $\Ket{0}$, whose probability then could be measured by optical fluorescent detection: $P_0 = \cos^2{(\phi_g/2)} = \cos^2{(\frac{1}{32\hbar}gt_3 ^3 g_{NV}\mu_B \frac{\partial B}{\partial x}\cos{\theta})}$.  
Practically, the particle will be re-trapped for repeated measurement and either $\theta$ or $t_3$ would be used as a controllable parameter that shifts the value of $\phi_g$, with respect to which a fringe of $P_0$ is resolved.

 {\em Thermal state.---}Remarkably, the phase $\phi_g$ is independent on the initial motional condition, featuring an immunity to the initial motional noise in our scheme; consider an initial motional state $\rho_{th}=\int_{\beta} d^2\beta P_{th}(\beta)\Ket{\beta}\Bra{\beta}$, where $P_{th}$ is the Glauber P representation for thermal state.  The spin is initialised to $\Ket{0}$ in the trap so that it is decoupled from the motion, and as soon as the particle is released, $\Ket{0}$ is converted to ($\Ket{+1}+\Ket{-1})/\sqrt{2}$. Then at time $t= t_3$ we have, 
\begin{equation} 
\begin{aligned}
\rho_{th}(t_3)     =& \frac{1}{2} \int_{\beta} d\beta P_{th}(\beta) \Ket{\psi(t_3)}_{\beta}\Bra{\psi(t_3)}_{\beta}\\
                       &\otimes    (\Ket{+1} + e^{i\phi_g}\Ket{-1})  (\Bra{+1} + e^{-i\phi_g}\Bra{-1}).\\
\end{aligned}
\end{equation} 
Obviously the state of composite system is again factorizable (separable), so the phase difference accrued by the spin states is not affected by initial thermal motion. A feedback cooling on initial state of the c.m. to mK \cite{feedbackCooling} (by which the harmonic potential could barely sustain the thermal excitations) will suffice. This factorizability despite the untrapped motion (which naturally gives rise to dispersion) is a non-trivial feature of our scheme. 

{\it Experiment parameters.---}The maximum spatial separation $\Delta x_M$ of the superposed components is given at half the propagation time $t_3$ by 
\begin{equation}
\Delta x_M = 2 \times \frac{1}{2m}g_{NV}\mu_B \frac{\partial B}{\partial x} (t_3/4)^2,
\end{equation}
we now analyse the achievable scale of magnitude this separation under realistic parameters. We consider a diamond sphere of radius $R\sim 100$ nm and with a density of 3500 kg/m$^3$, whose mass is then $\sim 1.25 \times 10^{-17}$ kg (corresponding to $7.5 \times 10^{9}$ amu). 
The coherent evolution time $t_3$ is limited by the coherence time of the system accounting for all possible detrimental effects, which will be discussed below and here we suppose a realistic value of $\sim100\,\mu$s. Under an field gradient of $\frac{\partial B}{\partial x} \sim 10^{7}$ T/m \cite{MaminNatNano, Tsang_IEEE} we immediately obtain a separation of $\Delta x_M\sim 100$ nm. Interestingly, this is comparable to the size of the test nano object. So a good position measurement at time $t_3/2$, such as those used in feedback cooling \cite{feedbackCooling}, can even discriminate the two components of the superposition spatially. Of course, this measurement will destroy the superposition so that the superposition has to be tested through the $\phi_g$ induced fringes in other runs of the experiment where measurements are only done at $t_3$. Nonetheless some runs of the experiment measuring spatial position at time $t_3/2$ will confirm the picture that the components superposed are indeed spatially separated by 100 nm. 

\begin{figure}[h]
\centering
\includegraphics[width=6cm]{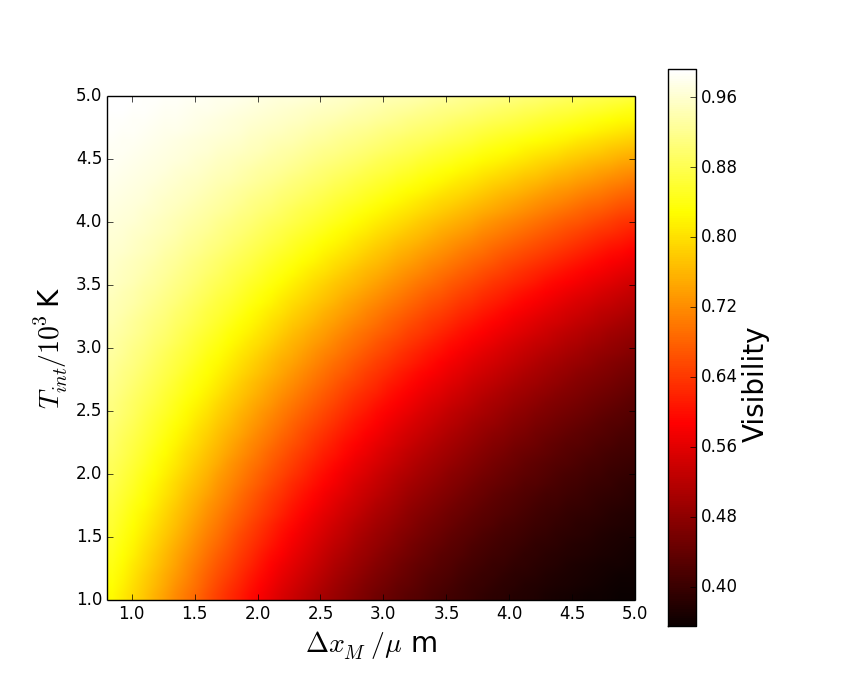}
\caption{Estimation on motional decoherence: $\Delta x_M$ is the maximum spatial separation and $T_{int}$ is the internal temperature of the test object. A large high visibility window indicates the strong robustness of our scheme against motional noise.}
\label{Visibility}
\end{figure}

{\it Decoherence.---}Collisional and thermal decoherence are mostly considered in matter wave interferometry and optomechanical systems \cite{TLDecoherence, LuclaNature2004}, which can be seen as random momentum kicks during the propagation of the matter wave and whose microscopic description is given by the master equation \cite{JamesNatureCom},
\begin{equation}
\mathcal{L}_i(\rho) = \int d\omega \gamma_{i}(\omega)\int_{|n|=1}\frac{dn^2}{4\pi}[e^{\frac{i\omega n_x}{c}\hat{x}}\rho e^{\frac{-i\omega n_x}{c}\hat{x}} - \rho],
\end{equation}
where $i$ indicates the specific decoherence class, including collisions with residual gas partials, scattering and absorption of blackbody photons, and thermal emission of radiation. $\gamma_{i}$ is the spectral rate and  {\bf n} is the direction cosine of the random momentum kick. Given $\gamma_{i}$ from realistic data the above master equation could be numerically simulated together with the unitary part of the free propagation (acceleration). 
Due to the entanglement between the spin and mechanical states, the motional decoherence process, specifically the part of which that carries out the which-path information of the two counter propagated wave packets, would eliminate the coherence of the c.m. and the reduced spin system at the end, which subsequently reduces the visibility of the following Ramsey measurement. 
Practically, the collisional decoherence is suppressed by preparing the system in a high vacuum chamber. As trapping is lifted during the flight the photonic scattering is absent, leaving only the radiative decoherence from background and black body radiation of the particle \cite{JamesNatureCom}. Here we provide a theoretical estimation of the upper bound of the detrimental effect from radiative decoherence by considering the worst scenario in the evolution \cite{supplementary}. The resultant interferometric visibility (square modulus of the off-diagonal term of the reduced density matrix of the spin system) of the Ramsey measurement is shown in Fig. \ref{Visibility}.  \\
Spin dephasing of the NV center will be the last detrimental effect that limits the absolute coherence time of the system. NV centers in isotopically-purified bulk diamond can have electron spin coherence time $T_2$ up to $\sim 2 $ ms at room temperature\cite{balasubramanian2009ultralong}, but such exceptional times have not been found in nanodiamonds. In order to achieve the longest $T_2$, nanodiamonds are made from high purity bulk material with a low density of nitrogen impurities and $^{13}$C. Nanodiamond pillars with 300-500 nm diameter have shown a spin echo $T_2$ time of over 300 $\mu s$ \cite{andrich2014engineered}. Pillars with 50 nm diameter and 150 nm length have achieved a spin echo $T_2$ time of $79$ $\mu s$ \cite{trusheim2013scalable}. This time was further extended by appropriate decoupling techniques. Interestingly, as an additional advantage, the sequence of MW pulses applied in our scheme, namely $(\pi/2)_x-(\pi)_x-(\pi)_x-(\pi/2)_x$, is a dynamical decoupling sequence \cite{CPMG} that would echo out the noise attributed to any slow and  spin relevant effect such as quasi static spin bath. More interestingly the perturbation from the rotational DOF, induced by an unknown torque on the particle since NV center is not necessarily situated at the c.m. of the nanodiamond, could also be suppressed by virtue of this technique \cite{supplementary}. 
{\it Testing spontaneous collapse models.---}Using the macroscopicity measure in Ref.\cite{nimmrichter2013macroscopicity}, a high visibility of our interferometry would impose a value of $\mu = 24$ for our system, which is comparable to the largest among the proposed experiments to date, such as those employing oscillating micromirrors or larger molecules. Since macroscopicity is intimately connected to the testability any macrorealistic modification of quantum theory, in this regard, another key purpose for the creation of the spatially large superposition will be to test the continuous spontaneous localiztion (CSL) model \cite{CSLPearle1989, NonmarkovianCSL, AlderCSL}, which is characterized by the locallization length $r_{CSL}$ and rate $\lambda_{CSL}$. The former is about $100$ nm that sets the scale above which the delocalized matter wave gets localized. The latter represents the average collapse rate at one proton mass, on which the interferometric experiment could place a bound. For our scheme, if we were to observe a high visibility (as expected from the above considerations of environmental decoherence), it would bound the collapse rate to \cite{bassi2013models}
\begin{equation} 
\lambda \leq 1/2N^2 t_3 \sim 10^{-14} \quad s^{-1},
\end{equation} 
where N is number of protons of our test object, which is $10^9$ in our case. The version of CSL by Adler ($\lambda \sim 10^{-9}$ $s^{-1}$) \cite{AlderCSL} should thus already decohere our superposition by a mechanism beyond standard quantum theory, while, to access the version by GRW \cite{GRW1986} ($\lambda \sim 10^{-16}$ $ s^{-1}$) one will need to extend the coherence time of the NV centre spin by two orders of magnitude, which is challenging.

{\em Other intrinsic decoherence.---}In order to unambiguously test CSL, it is crucial to rule out the significance of other hypothetical localization effects in the mescoscopic region we are considering. For instance, the gravitational time dilation effect \cite{IgorNP}, which couples the internal degree of freedom to the c.m. motion of a compound system when the state of the latter is spatially separated in the direction of a gravitational field, will induce a dephasing process on a c.m. subsystem. Substituting the relevant parameters of our model ($T_{int} = 400$ K, $\Delta x = 100$ nm and $N = 10^{9}$) we immediately obtain a coherence time admitted by this time dilation effect of 1000 s, which is sufficiently far from the scale of the coherence time we consider. In a similar vein, if we consider gravitational reduction models \cite{Penrose}, then, assuming mass density concentrated around nuclei \cite{Kleckner}, we obtain a decoherence time of $100$ s. Moreover, by engineering
a superposition of distinct kinetic energy states by changing the initial spin state to $(\Ket{0} + \Ket{+1})/\sqrt{2}$ in our free-flight scheme, we can constrain an effective parameter $\Theta$ of space-time textures \cite{Hu-Anastopoulos} to $\lesssim 10^{25}$ contingent on a high interferometric visibility.

{\em Multiple NVs.---}Diamond samples with multi-NVs are easy to obtain and provide a large spin-dependent fluorescence increasing the sensitivity of the final spin measurement.  It has been experimentally demonstrated that the orientations of all those NV centers' axes could be identically alligned to one of the four possible directions in the diamond crystal and their spin states could also be collectively manipulated and measured with Ramsey pulses \cite{EdmondsPRB2012, LesikAPL2014, MichlAPL2014}. The mechanism in this multi NVs scenario will follow the similar formula developed above \cite{supplementary}, starting with an {\em arbitrary} coherent state for the c.m. and a $l$ fold product state of $(\Ket{+1} + \Ket{-1})/\sqrt{2}$ for spin ensemble, the composite system ends up again a separable state in which the spin state is trivially a $l$ fold product state of $(\Ket{+1} + e^{-i\phi_g} \Ket{-1})/\sqrt{2}$. Evidencing this accrued phase on the multi-spin ensemble would reveal the superposition of the intermediate state of the corresponding collective spin-c.m. system.
 

{\em Conclusions.---}We have shown a method to generate and evidence superpositions of two c.m. states of a free (in the sense of being untrapped) nano-object of $\sim 10^9$ amu mass. The untrapped nature of the particle, in conjunction with spin dependent acceleration/deceleration in an external magnetic field gradient enables us to reach 100 nm spatial separations between the superposed components. This can open up possibilities of testing some of the spontaneous collapse models such as Adler's model \cite{AlderCSL} through a method that is qualitatively very different from the recently proposed non-interferometric tests \cite{optomech_testCSL}. The scheme completely surpasses the scale of the spatial separation possible through a trapped particle of the same mass \cite{scala2013matter} by $10^3$ orders of magnitude (essentially due to the absence of a finite frequency). In comparison to the adaptation of the Ramsey Bord\'e technique to nano-objects \cite{Plenio2014QGtest}, we have employed a state-dependent force that significantly boosts the delocalization scale of the matter wave. Such a macroscopicity is unattainable via photonic momentum kicks in Ramsey Bord\'e method, and the concomitant Doppler dephasing  is negligible in our NV case \cite{supplementary}.
A positive feature is that the relevant interferometric phase can be probed solely via spin Ramsey interferometry without directly measuring matter wave distribution \cite{CSLtestTalbotlau, JamesNatureCom}. Moreover, from the point of view of control, an electron spin in solid is a promising system with lower noise compared to optical frequency fields in cavity-optomechanics, while its coupling to the c.m. through a magnetic field gradient could potentially be easier than achieving strong optomechanical couplings.  Uniquely, the MW control is also naturally a dynamical decoupling that suppress those slow detrimental dynamics, so that the best coherence times of $100$ $ \mu$s can be used. The fact that the scale of spatial separation can be increased substantially by using untrapped particles, and yet be evidenced solely by a spin-only Ramsey interferometry in a gravitational potential, and indeed be independent of both the initial thermal state of the nano-object and its mass greatly facilitates the possibility of the interferometric probing of large superpositions. In view of the fact that manipulation of a spin-full levitated nano-object is being intensely pursued experimentally \cite{CoolingLevitatedObject} at the moment, our scheme should be realizable in the near future.

\section*{Acknowledgments}
We acknowledge the EPSRC grant EP/J014664/1. This work was also supported by EPSRC as part of the
UK Hub in Networked Quantum Information Technologies (NQIT), grant EP/M013243/1. GWM is supported by the Royal Society. HU acknowledges support by the John F Templeton foundation (grant 39530) and the Foundational Questions Institute (FQXi). C. Wan is sponsored by Imperial CSC scholarship.

\section*{Supplementary Material}
In this supplementary material we provide details on the calculation and simulation of the dynamics considered in the main text.  Further discussion on the noisy reduction and spin manipulation sequence are presented in the following sections. 
\section{Evolution of the free particle under spin dependent force}
The unitary evolution of the scheme we discussed above can be represented as:
\begin{equation}
\begin{aligned}
U_{tot} = &U(\tau_3) \cdot C\cdot U(\tau_2)\cdot C \cdot U(\tau_1)\\
=&e^{-iH\tau_3/\hbar}\cdot C\cdot e^{-iH\tau_2/\hbar}\cdot C\cdot e^{-iH\tau_1/\hbar},
\end{aligned}
\end{equation}
where $C$ is the spin flip operator, whose exact realisation will be discussed in next part, and ideally we assume the spin flip is done instantaneously, operationally represented by $C = (\Ket{-1}\Bra{+1} + \Ket{+1}\Bra{-1})$. The time intervals we have are $\tau_3 = t_3 - t_2$, $ \tau_2= t_2 - t_1$ and $ \tau_1 = t_1$. On could re-arrange the above expression in a operator-wise way that
\begin{widetext}
\begin{equation}
\begin{aligned}                                     
               U_{tot}= & e^{\frac{i}{3m}(\sum_i^3 \tau_i^3F_i^2-\tau_2^2\tau_3F_2^2-\tau_1^2(\tau_2+\tau_3)F_1^2)-\frac{i}{3m}(\tau_1\tau_3^2F_1F_3+\tau_1\tau_2^2F_1F_2+\tau_2t_3^2F_2F_3+2\tau_1\tau_2\tau_3F_1F_2)}\\
                & \times e^{i\hat{x}\sum_i^3F_i\tau_i} \times e^{\frac{-i\sum_i^3t_i}{m}\hat{p}^2}e^{-\frac{i}{2m}\hat{p}(\sum_i^3 \tau_i^2 F_i + 2\tau_2\tau_3F_2 + 2\tau_1(\tau_2+\tau_3)F_1)},\\
\end{aligned}\label{FINAL_UNI}
\end{equation} 
\end{widetext}
where $F_1 =F_3= \mu_Bg_{NV}\frac{\partial B}{\partial x} \hat{S_z}-mg\cos{\theta}$, and $F_2 = -\mu_Bg_{NV}\frac{\partial B}{\partial x} \hat{S_z}-mg\cos{\theta}$. Given by $\tau_1=\tau_3=\tau_2/2 $ as required separable condition, we have
\begin{equation}
U_{tot} = e^{\frac{-t_3^3(\mu_Bg_{NV}\partial B/\partial x \hat{S_z}-mg\cos{\theta})^2}{24m}}e^{\frac{-it_3}{m}\hat{p}^2},
\end{equation}
which is simply a wave packet spread operator plus a spin dependent phase shifting term.  Now apply this unitary operator to our initial state $\Ket{\psi(0)}=\Ket{\beta}(\Ket{+1} + \Ket{-1})/\sqrt{2}$ and solve it in the position representation, giving (by dropping any global phase terms)
\begin{widetext}
\begin{equation}
\braket{x|\Psi(t_3)} = e^{-ip_0x/\hbar} e^{-\frac{(x-x_0-p_0t_3/m-g\cos{\theta}t_3^2/2)^2}{\frac{4\hbar}{m\omega}(1+(\omega t_3)^2/16)}}
\times \frac{1}{\sqrt{2}}(\Ket{+1} + e^{-i\mu_Bg_{NV}\partial B/\partial x t_3^3\cos{\theta}g/16\hbar}\Ket{-1}),
\end{equation}
\end{widetext}
where $\omega$ is the trapping frequency before the particle is released, $p_0$ and $x_0$ is the initial position and momentum of the partial, respectively. Clearly the phase information accumulated on spin state is neither dependent on the initial kinetic condition of the trapped particle (for instance $p_0$ or $x_0$), nor any disturbance on the frequency $\omega$ of the trap. 

\section{Spin flip pulse sequences }
\begin{figure}[h]
\centering
\includegraphics[width=6cm]{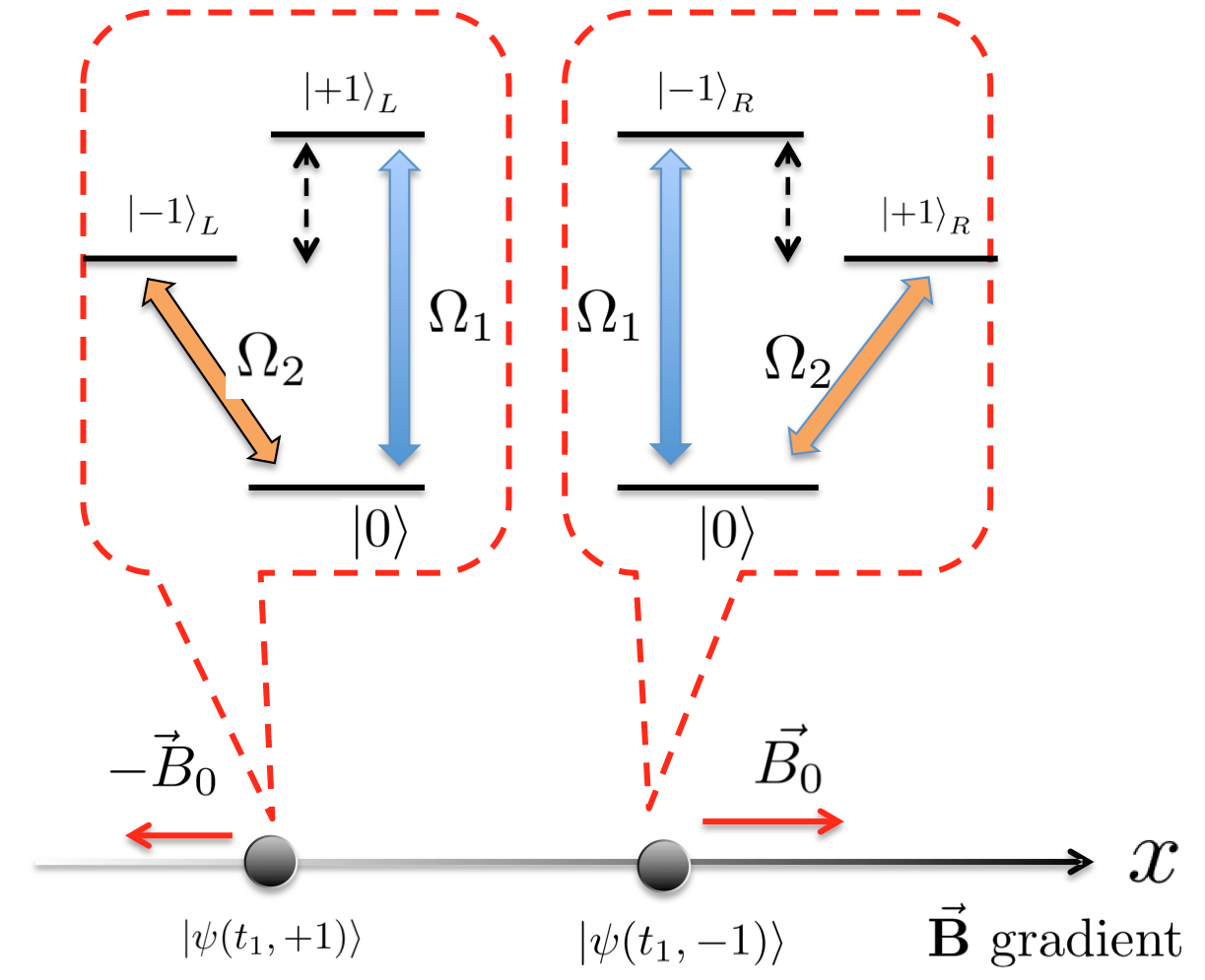}
\caption{Position dependent energy split of the spin system due to the delocalized CM state of the nano diamond in an inhomogeneous magnetic field}
\label{Spin_diagram}
\end{figure}

The direct transition from spin $\Ket{+1}$ to $\Ket{-1}$ is not dipole allowed since two quanta of angular momentum would be required, so here we propose a multi-MW pulses sequence that could introduce a coherent transition between $\Ket{+1}$ and $\Ket{-1}$ mediated by spin $\Ket{0}$ state. As discussed above, the spin flips happen at $t_1$ when the counter-propagating components are located at $x=\Delta x_M/4$ and $x=-\Delta x_M/4$, respectively. The local magnetic fields seen by the NV center have the same magnitude but opposite directions, therefore the corresponding energy diagram for the spin system has the  same Zeeman splitting between $\Ket{+1}$ and $\Ket{-1}$, but their ordering is swapped, as shown in Fig. \ref{Spin_diagram}. 
Now suppose we want to flip the system state from $(\Ket{\psi(t_1, +1)}\Ket{+1} + \Ket{\psi(t_1, -1)}\Ket{-1})/\sqrt{2}$ to $(\Ket{\psi(t_1, +1)}\Ket{-1} + \Ket{\psi(t_1, -1)}\Ket{+1})/\sqrt{2}$. The first MW pulse is sent to drive a Rabi oscillation $\Omega_1$ between ground state $\Ket{0}$ and the degenerate states of $\Ket{+1}_L$ and $\Ket{-1}_R$, without affecting the other level and after a certain duration of the pulse it brings the spin state to $\Ket{0}$. This is followed by a second MW pulse that hits the transition between $\Ket{0}$ and the degenerate states of $\Ket{-1}_L$ and $\Ket{+1}_R$ fully transferring population to these latter states.
Therefore we could effectively reverse the spin state in each superposed state by a two MW pulse sequence. The shortest pulse duration of MW accessible would be of $10$ ns, corresponding to a bandwidth of $\sim100$ MHz.  The Zeeman splitting between the $\Ket{+1}$ and $\Ket{-1}$ spin states is about 56 GHz/T $\times \partial B/\partial x \times \Delta x_M/4 \sim 56$ GHz (assuming a $10^7$T/m magnetic inhomogeneity), which is sufficiently large to resolve the two resonance $\Omega_1 $ and $\Omega_2$. 

\section{the torque echo effect}
The MW pulse sequences in our protocol is naturally a dynamical decoupling scheme, which is ubiquitously in used for suppress the environmental noise with a relative low frequency (comparing with the evolution of the central system). In our case the type Ib diamond sample in used has a rich nitrogen impurities which has been widely observed as a quasi static spin bath inn bulk diamond sample, and whose noise effect on the NV center is effectively a magnetic field fluctuation. 

Moreover, as the free particle our nanodiamond possesses rotational degree of freedom and due to the fact that the NV center is not situated at center of mass of the particle, inevitably there would be a spin relevant torque on it that may affect the translational decoherence via spin coupling. Presumably this rotation is slow with respect to the translational evolution, our MW decoupling sequence would echo out such effect: The NV position in diamond is indeed unknown to us so the $\Ket{+1}$ spin state will experience a unknown torque in some direction and so does the $\Ket{-1}$ with the same magnitude but opposite direction. When we slip the $\Ket{+1}$ and $\Ket{-1}$ state (the $\pi$ pulse in the middle of the evolution), the superposition states of the diamond will then begin to experience the same unknown torque but in the other direction. As the two periods (for which the spin state being $\Ket{+1}$ or $\Ket{-1}$ and the other way) are equal, finally the two superposition components end up with no net torque. 

\section{Other systematic noise} 
The magnetic field generated by a fixed magnet will drift in a timescale of hours, which is much larger than the  system evolution time of $ 1$ ms. Therefore the related noise effect from the magnet could be neglected. Practically we should require the magnet sample to be "single domain" in order to avoid the effect of domain wall moving. 
The Doppler effect may appear in the spin state preparation when the nanoparticle is oscillating in the trap. We will show that such the effect is negligible for the case of NV spin under microwave driving. Here we consider a coherent state for the c.m. of the NV center.  The Doppler-shift linewidth could be found by (in an estimation)
\begin{equation}
\delta f = f_0 \times \frac{v_0}{c} = f_0 \times \frac{z_0\omega_z}{c},
\end{equation}
where $f_0 = 2.87$ GHz is the microwave frequency in use and $c$ is the speed of light. $v_0$ and $z_0$ are the velocity and displacement amplitude of the harmonic oscillation, respectively. We consider that the nano-object traverses about $z_0 \approx 100$ nm in the $100 \mu$s experimental time-scale, which gives an average $v_0 \approx 0.001$ m/s. We then eventually have $\delta f = 0.029$ Hz. This is much smaller than the typical linedwidth of an NV center of $\sim10$ MHz \cite{Linewidth_NV}. For a thermal state of the CM, with a temperature cooled to about 1mK in the trap, the root-mean-square velocity is about $v_1= \sqrt{3kT/m}\sim 0.002$ m/s therefore the Doppler shift would not be a concern in our scheme.\\

\section{Estimated bound for motional decoherence}
Practically in order to keep the decoherence process sufficiently low, we suppress the collisional process by preparing the system in a high vacuum environment. Radiative decoherence is determined by the complex refractive index at typical wavelengths of room temperature blackbody radiation and the internal temperature of the test object, which had been unavoidably heated up \cite{JamesNatureCom} due to the initial optical trap.
The full dynamics of the free flight scheme is described by the master equation, 
\begin{equation}
\dot{\rho} =-\frac{i}{\hbar} [H_s, \rho] + \sum_i\mathcal{L}_i(\rho),
\label{MasterE}
\end{equation} 
where $H_s$ is the Hamiltonian (1). Starting with an initial state $\Ket{\psi{(0)}} = \Ket{\beta}\otimes(\Ket{+1} + \Ket{-1})/2$, we want to estimate the decoherence effect accumulated on the intermediate state (4) during the time evolution. Due to the isotropic and perturbative nature of the noise considered(momentum kicks), the maximum spatial separation of the counter propagating wave packets remains unchanged, hence only a dephasing on CM state is involved. Note that in the unitary case the spread of the Gaussian packets of each of the two superposed components is up to 10 times the initial ground state width $10\sigma_0 = 10 \sqrt{\hbar/m\omega} \sim 0.1$ nm, which is much smaller than the maximum spatial separation between them $\Delta x_M \sim 100$ nm (taking the initial trap frequency $\omega$ as $10^5$ Hz). Therefore an estimation on the upper bound of the motional decoherence effect could be obtained as followed: we simulate the coherence loss under Hamiltonian (1) with our composite system being in a virtual state, 
\begin{equation}
\Ket{\Psi}^{\prime}= \frac{\Ket{+\Delta x_M/2}\Ket{+1} + \Ket{-\Delta x_M/2}\Ket{-1}}{\sqrt{2}}
\end{equation}
where $\Ket{\pm\Delta x_M/2}$ is the eigenstate of position projection operator, $\hat{x} \Ket{\pm \Delta x_M/2}=\pm\Delta x_M/2\Ket{\pm\Delta x_M/2}$ and taking $H_s =1$. This state would suffer the maximum dephasing effect that would happen in the free flight scenario we consider above (namely the evolution starting with $\psi(0)$ under full Hamiltonian (1)), which could overestimate the possible decoherence, hence impose an upper bound on it. 
At time $t$ the off diagonal term of the reduced spin system, which directly links to the visibility of the Ramsey interferometry is 
$\Bra{+1} \textrm{Tr}\{ \rho(t)\}\Ket{-1}
=e^{-\eta(\Delta x_M)t} $
where 
\begin{equation}
\eta(\Delta x_M) = \sum_i\int d\omega \gamma_{i}(\omega)\int_{|{\bf n}|=1}\frac{dn^2}{4\pi}[e^{\frac{i\omega n_x}{c}\Delta x_M} - 1],
\end{equation}
$\rho(t)$ the system density operator and the trace operation is carried over the motional degree of the CM.  The numerical result is obtained by sampling the spectrum of the diamond response and the blackbody spectrum for a range of internal temperatures, as shown in Fig.[2].

\section{Multi-NVs scenario}
For a nano diamond with $l$ NVs (realistically $l$ is of the order of 100s), the initialisation of the spin system (after the first  MW pulse on the nano diamond) would be a product state of those individual spin systems: $\Ket{\psi}_{spin}^l = (\Ket{+1} + \Ket{-1})^{\otimes l}/2^l$,  to which the CM motion couples magnetically, they are coupled under new Hamiltonian,
\begin{equation}
H = \frac{\hat{p}^2}{2m} -  \sum_{i=1}^l g_{NV}\mu_B \frac{\partial B}{\partial x}\hat{S_z}^{(i)}  \hat{x} + m g\cos{\theta} \hat{x}.
\label{Hamil_sys_M}
\end{equation}


Note that since only the spin state $\Ket{\pm1}$ of each NV centre is relevant to the ballistic expansion process of the matter wave, we could describe each spin as a pseudo-spin-1/2 system and characterise the output state by a collective spin ${\bf J}$ summing up all the pseudospins. $\Ket{\psi}_{spin}^l$ could be rewritten as 
\begin{equation}
\Ket{\psi}_{spin}^l = \frac{1}{2^l} \sum_{n=0}^l \frac{l!}{n!(l-n)!} \sum_k\Ket{+1}^{\otimes n} \Ket{-1}^{\otimes (l-n)},
\end{equation}
which could be seen as a superposition of N-qubit symmetric Dicke state:
$\Ket{D_l^n} =  \sum_k\Ket{+1}^{\otimes n} \Ket{-1}^{\otimes (l-n)}$,
where the summation $k$ is over all the permutations of $n$'s $+1$ and $(l-n)$'s $-1$. During the free propagation,  the matter wave of the nano object would undergo a ballistic expansion, whose trajectories are determined by the spin value $m=2n-l$ of the corresponding Dicke state. The intermediate state at time $t$ in this scenario is
$
\Psi_{sys}(t) =  \sum_n \Ket{\psi, m} \Ket{D_l^n}/2^l
$,
where $\Ket{\psi, m}$ is the motional state of the CM that coupled to Dicke state $\Ket{D_l^n}$ with a spin value of $m$. Interestingly, at time $t_3$ the final state of the composite system is again separable,
\begin{equation}
\Psi_{sys}(t_3) = \frac{1}{2^l}\Ket{\psi(t_3)}  \sum_n e^{i(2n-m)\phi_g}\Ket{D_l^n},
\label{Multi_composite}
\end{equation}
where $\Ket{\psi(t_3)}$ has the formula of Equ. (4) in the main text in $x$ representation. 
 The reduced spin state of (\ref{Multi_composite}) is exactly an $l$ product state of $(\Ket{+1} + e^{i\phi_g} \Ket{-1})/\sqrt{2}$, on the basis of which this phase could be measured via the Ramsey pulse and hence reveal the intermediate superposition $\sum_n \Ket{\psi, m} \Ket{D_l^n}/2^l$.\\

\end{document}